\documentclass[icps3]{svjour}
\usepackage{graphicx}
\begin{document}

\title{Quantum Hall Stripe States in a Tilted Magnetic Field}


\author{T. Jungwirth\inst{1,2,3},
  A.H. MacDonald\inst{1,3}, 
  L. Smr\v{c}ka\inst{2}, 
  S.M. Girvin\inst{1}}

\authorrunning{T. Jungwirth et al.}

\institute{Department of Physics, Indiana University,
  Bloomington, IN 47405, USA
  \email{jungw@gibbs.physics.indiana.edu}
  \and 
  Institute of Physics ASCR, Cukrovarnicka 10, 162 00 Praha 6, Czech Republic
  \and
  Department of Physics, The University of Texas at Austin,
  Austin, TX 78712, USA}

\maketitle

\begin{abstract}
A strong anisotropy in the longitudinal resistivity of
a 2D electron system has been  observed at  half-filled high Landau levels.
We report on detailed Hartree-Fock calculations of the unidirectional
charge density wave (UCDW) orientation
energy induced by a tilted magnetic field. 
We find that stripes can orient both parallel or perpendicular to the
in-plane field depending on the sample parameters and field strength.
The close agreement between complex experimental data on different sample
geometries and our theoretical results strongly support the UCDW picture
as the origin of the observed anisotropies in high Landau levels.
\end{abstract}

\section{Introduction}

Recently, a strongly
anisotropic transport has been observed
in high
quality GaAs/Al$_x$Ga$_{1-x}$As single heterojunctions
\cite{horst:aps93,lilly:prl99a,du:ssc99,pan:prl99,lilly:prl99b}
at  filling factors
$\nu$ = 9/2, 11/2, etc., and in 2D hole
systems \cite{shayegan:physicae99}
starting at $\nu=5/2$.
The origin of the magnetotransport anisotropy has not been
firmly
established yet. The most appealing interpretation suggests
that the 2D electron gas
spontaneously breaks the translational symmetry by forming
a unidirectional charge density wave, as predicted by
Hartree-Fock theory
\cite{koulakov:prl96,moessner:prb96}.
Because of uncertainty about the reliability of this Hartree-Fock
prediction,
a special emphasis
is placed on tests of its ability to explain
experimental results on stripe orientation in tilted
magnetic fields.
 
Many-body RPA/Hartree-Fock calculations using a model, parabolic quantum well 
system showed \cite{jungwirth:prb99} 
that the stripe orientation relative to the in-plane field
is not universal. In narrow quantum wells, stripes are expected
to orient perpendicular to the in-plane field. For wider quantum wells, 
with two occupied subbands at zero tilt angle, 
a more complex behavior is predicted. Isotropic 2D electron systems
with half-filled lowest Landau level (LL) of the second subband can undergo a 
transition into a UCDW state due to level 
(anti)crossings induced by the in-plane magnetic field. The orientation
of these stripe states can change from parallel to perpendicular as
the strength of the in-plane field increases.

In this paper we present calculations of the field-tilt UCDW anisotropy energy
that include a
self-consistent
local-spin-density-approximation (LSDA) description of
one-particle
states in experimental sample geometries \cite{jungwirth:prb99}. 
Our results are compared
to magnetotransport experiments on GaAs/Al$_x$Ga$_{1-x}$As 
single heterojunctions with different 2D electron gas densities 
\cite{pan:prl99,lilly:prl99b,lilly:unpubl} and on
a square quantum well sample with two occupied electric subbands
\cite{pan:prl00}.
\section{Theoretical model and results}

The energy per electron of the UCDW state
at fractional filling $\nu^*$ of the valence LL
is given by \cite{koulakov:prl96} 
\begin{equation}
\label{ucdwen}
E=\frac{1}{2\nu^*}\sum_{n=-\infty}^{\infty} \Delta_n^2
U\left(\frac{2\pi n}{a}\hat{e}\right)\; ,
\Delta_n=\nu^*\frac{\sin(n\nu^*\pi)}{n\nu^*\pi}\; ,
\end{equation}
where $a$ is the period of the UCDW state, $\hat e$ is the
direction of charge variation, and
$\Delta_n$  is the Fourier transform of the the guiding
center occupation function at wave vector $ n 2 \pi /a$.
In Hartree-Fock theory, the UCDW state energy depends only on
$a$ and $\hat e$ and the optimal UCDW is obtained by
minimizing Eq.(\ref{ucdwen}) with respect to these parameters.
In Eq.(\ref{ucdwen}), $U(\vec{q})$
can be separated into  direct, $H(\vec{q})$,
and exchange,
$X(\vec{q})$, contributions with
\begin{eqnarray}
H(\vec{q})&=&\frac{1}{2\pi\ell^2}\, 
V(\vec{q})\nonumber \\
X(\vec{q})&=&-\int\frac{d^2p}{(2\pi)^2}\,
e^{i(p_xq_y-p_yq_x)\ell^2}\, V(\vec{p})\; ,
\label{hartree}
\end{eqnarray}
where $V(\vec{q})$ is the RPA-screened effective Coulomb interaction 
\cite{jungwirth:prb99}.

To evaluate $V(\vec{q})$
we choose the in-plane component $B_{ip}$
of the magnetic field to be in the
$\hat{x}$-direction and use the following Landau gauge for the vector
potential, $\vec{A}=(0,B_{\perp}x-B_{ip}z,0)$.
The one-particle orbitals 
are calculated using the self-consistent
LSDA method and for any sample geometry, with the $z$-dependent
confining potential, can  be written as
$e^{iky}/\sqrt{L_y}\,
\varphi_{i,\sigma}(x-\ell^2k,z)$,
where $k$ is the wave vector which labels states within LL $i$,
$\sigma$ is the spin index, and
$\ell^2 = \hbar c /e B_{\perp}$.  The translational symmetry responsible
for LL degeneracy leads to a 2D wavefunction
$\varphi_{i,\sigma}(x,z)$
which is independent of the state label $k$, except for the rigid shift
by $\ell^2k$ along $x$-axis.
This in turn leads to two-particle matrix elements of the
Coulomb interactions with a dependence on state labels which
is identical to that for the lowest
LL of a zero-thickness 2D electron system provided the
2D Coulomb interaction is replaced by
the  effective interaction 
$V(\vec{q})$. 

For an infinitely narrow electron layer the  orbital effects
of $B_{ip}$ vanish and the 2D Coulomb
interaction, $V(\vec{q})$, reduces to
$e^{-q^2\ell^2/2}/q$ $(L_N(q^2\ell^2/2))^2$
$2\pi e^2\ell/\epsilon$ where $L_N(x)$ is the
Laguerre polynomial and $N$ is the orbit radius quantum number.
Starting from $N=1$, zeros of $L_N(q^2\ell^2/2)$ occur at
finite $q=q^*$, producing a zero in the repulsive Hartree
interaction at  wave vectors where the attractive
exchange interaction is strong. For the half-filled valence
LL the
corresponding UCDW state
consists of alternating
occupied and empty stripes of electron guiding center states
with
a modulation period
$a\approx 2\pi/q^*$.

In finite-thickness 2D systems 
the dependence of the effective
interaction on wavevector magnitude $q$ and orientation $\phi$
relative to the in-plane field direction
can be accurately approximated \cite{jungwirth:prb99} by
$V(\vec{q})=V_0(q)+
V_2(q)$ $\cos(2 \phi)$. At weak $B_{ip}$,
the isotropic term $V_0(q)$
has a wavevector-dependence similar to that of the effective
interaction in
the infinitely narrow 2D layer.
The corresponding curve for the valence LL at
$\nu=9/2$ for a single heterojunction with 2D density $n_e=2.67\times
10^{11}$~cm$^{-2}$ \cite{lilly:prl99b} is shown in the lower inset of
Figure~1. 

The non-zero anisotropy coefficient $V_2(q)$ of the effective
interaction at $B_{ip}>0$, also shown in the inset,
is responsible for the
non-zero UCDW anisotropy energy $E_A$, defined
\cite{jungwirth:prb99} as the
total Hartree-Fock energy of stripes oriented parallel with
$B_{ip}$
minus the total energy of stripes  perpendicular to $B_{ip}$.
The main plot of Figure~1 shows $E_A$ calculated for single
heterojunction samples. The two curves correspond to high and
low density 2D electron systems with one occupied electric subband
in perpendicular magnetic field.
The  subband wavefunctions are plotted in the upper inset.
We found that stripes are oriented perpendicular
to the in-plane field for the considered interval of
densities $n_e=1.5-2.67\times
10^{11}$~cm$^{-2}$, consistent with available experimental data
on single heterojunctions  
\cite{pan:prl99,lilly:prl99b,lilly:unpubl}. Similar agreement
between theory and experiment is also abtained for other
half-filled high LLs.

In Figure~2 we present results for a 350~$\AA$ wide GaAs quantum well
with $n_e=4.6\times
10^{11}$~cm$^{-2}$ \cite{pan:prl00}. 
At $\nu=9/2$ and 11/2, and $B_{ip}=0$, the valence LL is the $N=0$
LL of the second subband. The effective interaction $V_0(q)$
has no zeros  at finite wavevector, as seen in the inset of Figure~2,
and the UCDW state is not expected to form, consistent with experiment
\cite{pan:prl00}. An in-plane field $B_{ip}\sim3$~T induces a level
anticrossing accompanied by a development of a  minimum in $V_0(q)$
at finite $q$.
As discussed above, 
it is the minimum of the interaction energy at
finite wavevector
that opens the possibility for the formation
of the UCDW state. The theoretical and experimental \cite{pan:prl00}
in-plane fields
corresponding to the onset of the UCDW and transport anisotropy, 
respectively,
are remarkably close.

The main graph in Figure~2 indicates that, in this sample, the sign of
$E_A$ changes at $B_{ip}\sim10$~T. The stripes orient
parallel with $B_{ip}$ at low  fields 
while perpendicular orientation
is obtained in the high in-plane field region,
consistent with
the experimentally observed interchange of easy and hard current axes
at similar $B_{ip}$.

We conclude that the close agreement 
between complex experimental data and theoretical
results for all studied samples leave little doubt as to the UCDW 
origin of the observed transport anisotropies.
\begin{figure}
\centerline{\includegraphics[width=8.5cm]{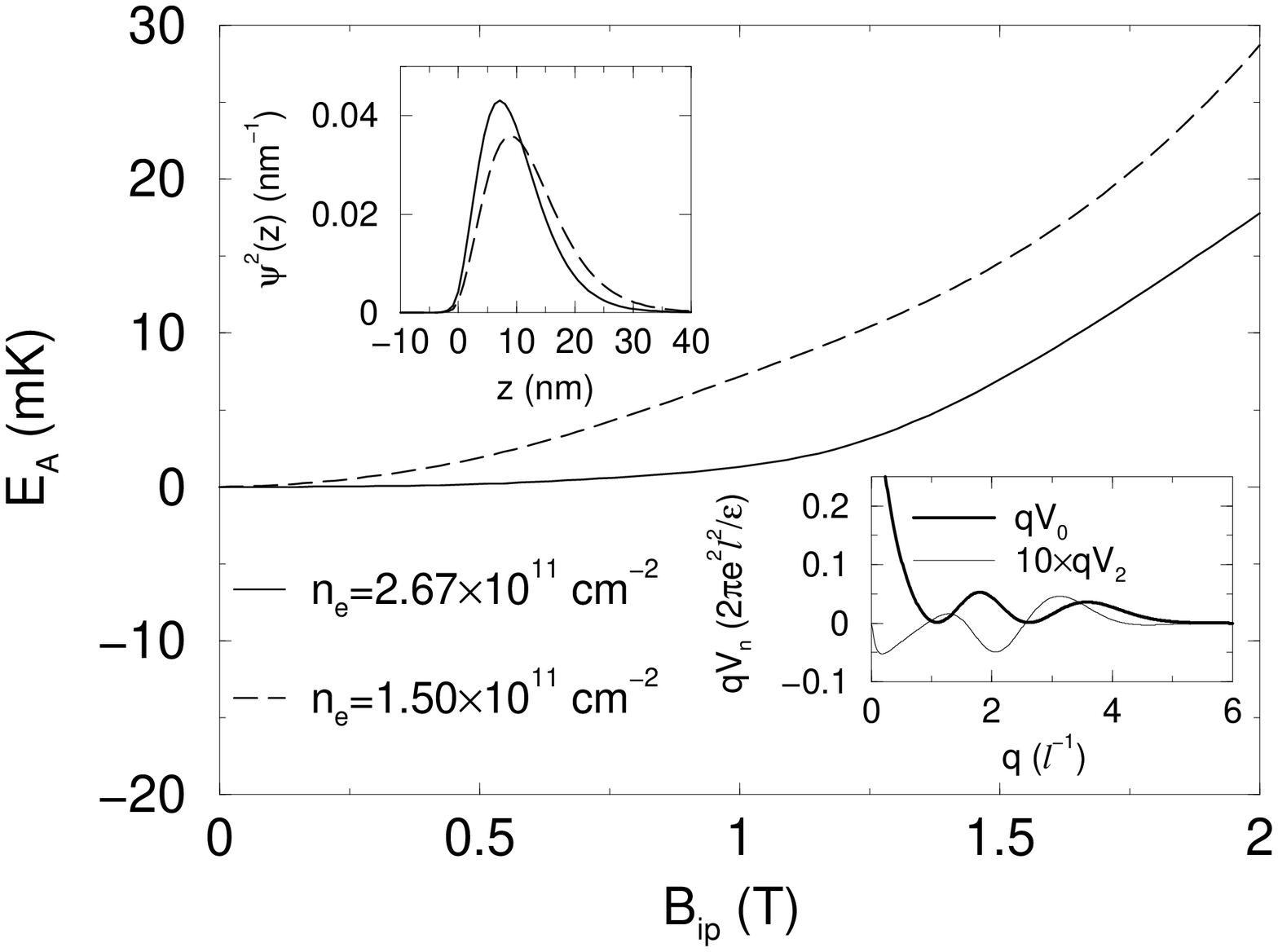}}
\caption{Main graph: UCDW anisotropy energy for single heterojunction
with 2D density $n_e=2.67$ (solid line) and 1.5 (dashed line) $\times
10^{11}$~cm$^{-2}$. Upper inset: square of the normalized subband wavefunction
for the two densities at $B_{ip}=0$. 
Lower inset: $n=0$ (thick line) and $n=2$ (thin line)
Coulomb interaction coeficients for
$n_e=2.67\times10^{11}$~cm$^{-2}$ and $B_{ip}=2$~T.
}
\label{fig1}
\end{figure}

\begin{figure}

\vspace{.5cm}

\centerline{\includegraphics[width=8cm]{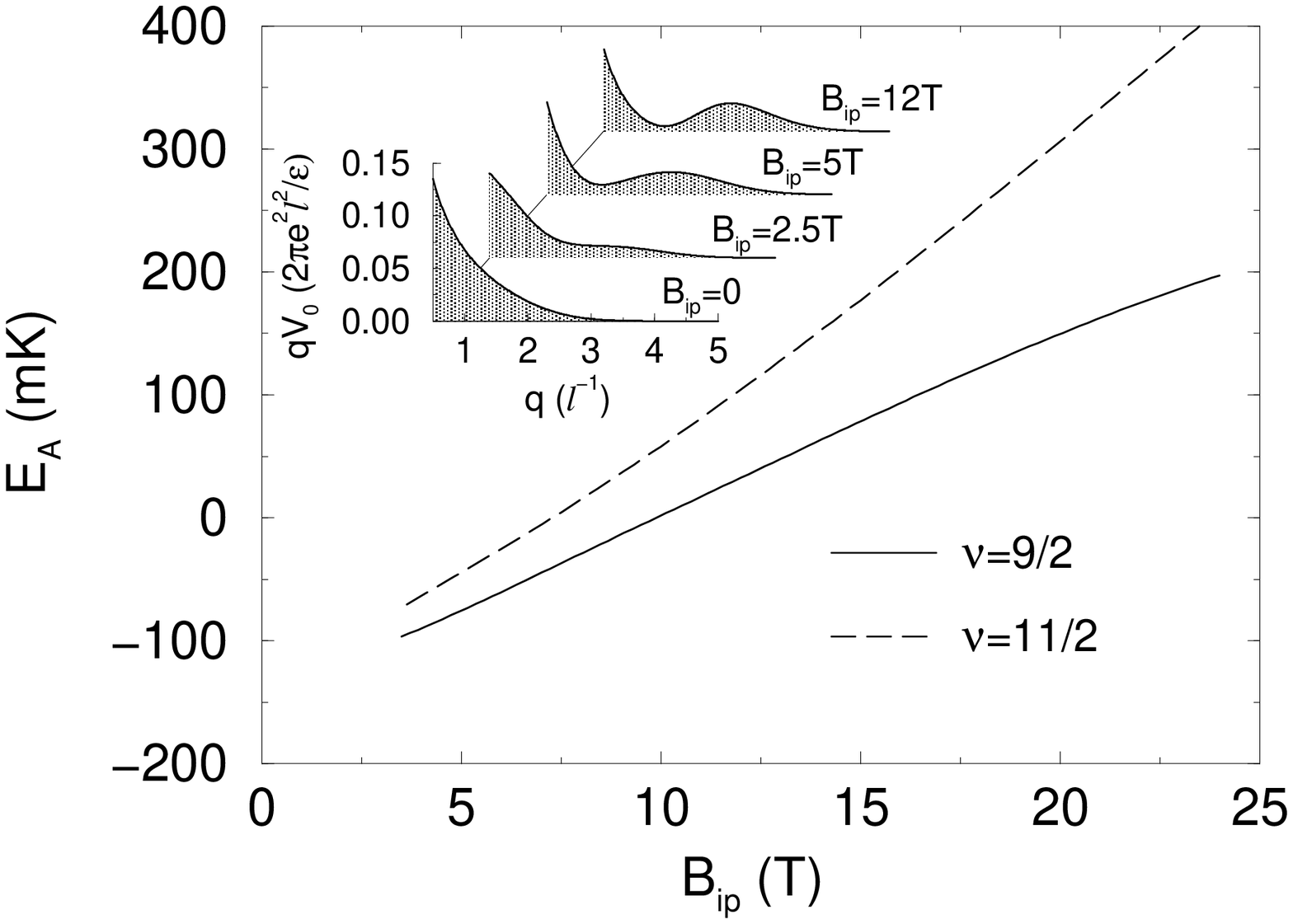}}
\caption{Main graph:  UCDW anisotropy energy for a square quantum well,
$n_e=4.6\times10^{11}$~cm$^{-2}$ and $\nu=9/2$ (solid line) and 
$\nu=11/2$ (dashed line). Inset: isotropic coefficient of the
 effective Coulomb interaction at $\nu=9/2$ and
different in-plane magnetic fields.
}
\label{fig2}
\end{figure}

\end{document}